\documentclass[aps,pra,twocolumn,amsmath,amssymb,showpacs]{revtex4-1}
\usepackage{txfonts}
\usepackage{epsfig,bm,dcolumn}
\usepackage{graphicx}
\usepackage{color}
\usepackage{overpic}

\begin{document}
\title{Anti-shadowing Effect on Charmonium Production at a Fixed-target Experiment Using LHC Beams}
\author{Kai Zhou$^{1,2}$, Zhengyu Chen$^1$, Pengfei Zhuang$^1$}
\address{$^1$ Physics Department, Tsinghua University and Collaborative Innovation Center of Quantum Matter, Beijing 100084, China}
\address{$^2$ Institute for Theoretical Physics, Johann Wolfgang Goethe-University Frankfurt, Max-von-Laue-Strasse 1, 60438 Frankfurt am Main, Germany}
\date{\today}
\begin{abstract}
We investigate charmonium production in Pb+Pb collisions at LHC beam energy $E_{\text {lab}}$=2.76 A TeV at fixed-target experiment ($\sqrt {s_{\text{NN}}}$=72 GeV). In the frame of a transport approach including cold and hot nuclear matter effects on charmonium evolution, we focus on the anti-shadowing effect on the nuclear modification factors $R_{AA}$ and $r_{AA}$ for the $J/\psi$ yield and transverse momentum. The yield is more suppressed at less forward rapidity ($y_\text{lab}\simeq$2) than that at very forward rapidity ($y_\text{lab}\simeq$4) due to the shadowing and anti-shadowing in different rapidity bins.
\end{abstract}
\pacs{25.75.-q, 12.38.Mh, 24.85.+p}
\maketitle

\section{Introduction}
\label{s1}

Recently a fixed-target experiment using the LHC beams has been proposed~\cite{brodsky}, where the study on quarkonia in nuclear collisions becomes specifically important, due to the wide  parton distributions in phase space which is helpful to reveal the charmonium production mechanism~\cite{lansberg}. Corresponding to the LHC beam energy $E_{\text {lab}}$=2.76 A TeV, where A is the nucleon number of the incident nucleus, the center-of-mass energy
$\sqrt{s_{\text{NN}}}$=72 GeV is in between the SPS and RHIC energies, and a quark-gluon plasma is expected to be created in the early stage of heavy ion collisions. Taking into account the
advantage of high luminosity in fixed-target experiments, which is helpful for detailed study of rare particles, the $J/\psi$ yield in Pb+Pb collisions at $E_{\text {lab}}$=2.76 A TeV per LHC
run year is about 100 times larger than the $J/\psi$ yield in Au+Au collisions at $\sqrt{s_{NN}}$=62.4 GeV per RHIC run year~\cite{brodsky}. With the high statistics, one may precisely distinguish between different cold and hot nuclear matter effects on charmonium production~\cite{andronic}. As is well known, the shadowing effect~\cite{vogt,eks}, namely the difference between the parton distributions in a nucleus and in a free nucleon, depends strongly on the parton momentum fraction $x$. Since $x$ runs in a wide region, $0.001\lesssim x \lesssim 0.5$, in the fixed-target experiments, it provides a chance to see clearly the shadowing effect on the charmonium distributions in different rapidity bins. In this paper, we study the shadowing effect on the nuclear modification factors for $J/\psi$ yield and transverse momentum in Pb+Pb collisions at LHC beam energy $E_{\text{lab}}$=2.76 A TeV.

\section{Evolution of Quark-gluon Plasma}
\label{s2}

The medium created in heavy ion collision at $\sqrt {s_{NN}}=72$ GeV is assumed to reach local equilibrium at a proper time $\tau_0$=0.6 fm/c~\cite{shen}, its consequent space-time evolution can be controlled by the ideal hydrodynamic equations,
\begin{eqnarray}
\label{hydro}
&& \partial_\mu T^{\mu\nu}=0, \nonumber\\
&& \partial_\mu j^{\mu}=0,
\end{eqnarray}
where $T_{\mu\nu}=(\epsilon+p)u_{\mu}u_{\nu}-g_{\mu\nu}p$, $j_{\mu}=nu_{\mu}$, $u_{\mu}$, $\epsilon$, $p$ and $n$ are respectively the energy-momentum tensor, baryon current, four-velocity of the fluid cell, energy density, pressure and baryon density of the system. The solution of the hydrodynamic equations provides the local
temperature $T(x)$, baryon chemical potential $\mu(x)$ and fluid velocity $u_\mu(x)$ of the medium which will be used in the
calculation of the charmonium suppression and regeneration rates~\cite{tang}. Taking the assumption of Hubble-like expansion and initial boost invariance along the colliding direction for high energy nuclear collisions, we can employ
the well tested 2+1 dimensional version of the hydrodynamics in describing the evolution of the medium created at $\sqrt {s_{NN}}=72$ GeV. Introducing the proper time $\tau=\sqrt{t^2-z^2}$ and space-time rapidity $\eta=1/2\ln\left[(t+z)/(t-z)\right]$
instead of the time $t$ and longitudinal coordinate $z$, the conservation equations can be simplified as~\cite{zhu}
\begin{eqnarray}
\label{hydro3}
&& \partial_{\tau}E+\nabla{\bf M} = -(E+p)/{\tau}, \nonumber\\
&& \partial_{\tau}M_x+\nabla(M_x{\bf v}) = -M_x/{\tau}-\partial_xp,\nonumber\\
&& \partial_{\tau}M_y+\nabla(M_y{\bf v}) = -M_y/{\tau}-\partial_yp,\nonumber\\
&& \partial_{\tau}R+\nabla(R{\bf v}) = -R/{\tau}
\end{eqnarray}
with the definitions $E=(\epsilon+p)\gamma^2-p$, ${\bf M}=(\epsilon+p)\gamma^2 {\bf v}$
and $R=\gamma n$, where ${\bf v}$ and $\gamma$ are the three-velocity of the fluid cell and Lorentz factor in the transverse plane.

To close the hydrodynamical equations one needs to know the equation
of state of the medium. From recent studies on particle elliptic flow and shear viscosity, the matter created in heavy ion collisions at RHIC and LHC energies is very close to a perfect fluid~\cite{song}. Considering that the momentum integrated particle yield, especially for heavy quarkonia, is not sensitive to the equation of state, we follow Ref.~\cite{sollfrank} where the
deconfined phase at high temperature is an ideal gas of gluons and massless
$u$ and $d$ quarks plus 150 MeV massed $s$ quarks, and the
hadron phase at low temperature is an ideal gas of all known
hadrons and resonances with mass up to 2 GeV~\cite{pdg}. There is
a first order phase transition between these two phases. In the
mixed phase, the Maxwell construction is used. The mean field
repulsion parameter and the bag constant are chosen as $K$=450
MeV fm$^3$ and $B^{1/4}$=236 MeV to obtain the
critical temperature $T_c=165$ MeV~\cite{sollfrank} at vanishing baryon number
density. Note that, when one calculates the rapidity or transverse momentum distribution of quarkonia, the choice of the equation of state may result in sizeable difference.

The initialization of the hot medium is taken as the same treatment in Ref.~\cite{zhu}. We use the final charged multiplicity to determine
the initial entropy density. For $\sqrt{s_{\text {NN}}}$=72 GeV, the charged multiplicity at central rapidity in center-of-mass frame is estimated to be
$dN_{\text {ch}}/d\eta=515$ based on the empirical formula~\cite{kestin}:
\begin{equation}
\frac{dN_{\text{ch}}}{d\eta}=312.5\log_{10}\sqrt{s_{\text{NN}}}-64.8.
\label{charged}
\end{equation}
The initial baryon density is obtained by adjusting the entropy per baryon to be 250~\cite{kolb}. From the empirical relation $\sigma_{NN}= 29.797+0.141(\ln\sqrt {s_{NN}})^{2.624}$~\cite{hikasa} between the inelastic nucleon-nucleon cross section $\sigma_{NN}$ in unit of mb and the colliding energy $\sqrt {s_{NN}}$ in unit of GeV, we have $\sigma_{NN}=36$ mb at $\sqrt {s_{NN}}$=72 GeV. These initial conditions lead to a maximum medium temperature $T_0$=310 MeV at the initial time $\tau_0$=0.6 fm/c. The medium maintains local chemical
and thermal equilibrium during the evolution. If we do not consider the charmonium interaction with the hadron gas, the charmonium distributions in the final state will be fixed at time $\tau_c$ corresponding to the critical temperature $T_c$ of the deconfinement phase transition.

\section{Charmonium Transport in Quark-gluon Plasma}
\label{s3}

Since a charmonium is so heavy, its equilibrium with the medium can
hardly be reached, we use a Boltzmann transport equation to describe
its phase space distribution function $f_\Psi(x,{\bf p}|{\bf b})$ in heavy ion collisions at impact parameter ${\bf b}$,
\begin{equation}
p^\mu \partial_\mu f_\Psi = - C_\Psi f_\Psi + D_\Psi,
\label{trans1}
\end{equation}
where the loss and gain terms $C_\Psi(x,{\bf p}|{\bf b})$ and $D_\Psi(x,{\bf p}|{\bf b})$ come from the charmonium dissociation
and regeneration in the created hot medium. We have neglected here the elastic scattering, since
the charmonium mass is much larger than the typical medium temperature. Considering that the feed-down from the excited
states $\psi'$ and $\chi_c$ to the ground state $J/\psi$~\cite{zoccoli} happens after the medium evolution, we should take transport equations for $\Psi=J/\psi,\ \psi'$ and $\chi_c$ when we
calculate the $J/\psi$ distribution $f_{J/\psi}$ in the final state.

Introducing the momentum rapidity
$y=1/2\ln\left[(E+p_z)/(E-p_z)\right]$ and transverse energy
$E_t=\sqrt {E^2-p_z^2}$ to replace the longitudinal momentum $p_z$ and energy $E=\sqrt{m^2+{\bf
p}^2}$, the transport equation can be rewritten as
\begin{equation}
\left[\cosh(y-\eta)\partial_\tau+{\sinh(y-\eta)\over \tau}\partial_
\eta+{\bf v}_t\cdot\nabla_t\right]f_\Psi=- \alpha_\Psi f_\Psi+\beta_\Psi
\label{trans2}
\end{equation}
with the dissociation and regeneration rates $\alpha_\Psi(x,{\bf
p}|{\bf b}) = C_\Psi(x,{\bf p}|{\bf b})/E_t$ and $\beta_\Psi(x,{\bf p}|{\bf b}) = D_\Psi(x,{\bf p}|{\bf b})/E_t$, where the third term in the square
bracket arises from the free streaming of $\Psi$ with transverse velocity ${\bf v}_t={\bf p}_t/E_t$ which leads to a strong leakage effect at SPS
energy~\cite{hufner}.

Considering the gluon dissociation $\Psi + g \to c+\bar c$ in the quark-gluon plasma, the dissociation rate $\alpha$ can be expressed as
\begin{equation}
\label{loss}
\alpha_\Psi=\frac{1}{2E_t}\int{d^3{\bf k}\over (2\pi)^3
2E_g}\sigma_{g\Psi}({\bf p},{\bf k},T)4F_{g\Psi}({\bf p},{\bf k})f_g({\bf k},T,u_\mu),
\end{equation}
where $E_g$ is the gluon energy, $F_{g\Psi}=\sqrt{(p k)^2-m_\Psi^2m_g^2}=p k$ the flux factor, and $f_g$ the gluon thermal distribution as a function
of the local temperature $T(x|{\bf b})$ and fluid velocity $u_\mu(x|{\bf b})$ determined by the hydrodynamics. The dissociation cross section in
vacuum $\sigma_{g\Psi}({\bf p},{\bf k},0)$ can be derived through the operator production expansion (OPE) method with a perturbative Coulomb wave
function~\cite{bhanot,arleo,oh,wang}. However, the method is no longer valid for loosely bound states at high temperature. To reasonably describe the
temperature dependence of the cross section, we take the geometric relation between the averaged charmonium size and the cross section,
\begin{equation}
\label{crosssection}
\sigma_{g\Psi}({\bf p},{\bf k},T)={\langle r^2\rangle_\Psi(T)\over \langle r^2\rangle_\Psi(0)}\sigma_{g\Psi}({\bf p},{\bf k},0).
\end{equation}
The averaged radial square $\langle r^2\rangle_\Psi(T)$ is calculated via potential model~\cite{satz} with lattice simulated heavy
quark potential~\cite{petreczky} at finite temperature. When $T$ approaches to the charmonium dissociation temperature $T_d$, the averaged radius square and in turn the cross section go
to infinity, which means a complete charmonium melting induced by color screening~\cite{matsui}. Using the internal energy $U$ as the heavy quark potential $V$, the dissociation temperautre
$T_d$ is calculated to be $2.1T_c, 1.16T_c$ and $1.12T_c$ for $J/\psi, \chi_c$ and $\psi'$, respectively~\cite{satz}.

The regeneration rate $\beta$ is connected to the dissociation rate $\alpha$ via
the detailed balance between the gluon dissociation process and its inverse
process~\cite{thews,yan}. To obtain the regeneration rate, we also need the charm quark distribution function in medium. Although the initially
produced charm quarks would carry high transverse momentum, they lose energy (momentum) when passing through the medium. Considering the experimentally
observed large open charm quench factor~\cite{star1,star2,alice1} and elliptic flow~\cite{phenix,alice3}, we take as a first approximation a
kinetically thermalized momentum spectrum for the charm quark distribution $f_c(x,{\bf q}|{\bf b})$. Neglecting the creation and annihilation of charm-anticharm pairs inside
the medium, the spacial density of charm quark number $\rho_c(x|{\bf b})=\int d^3{\bf q}/(2\pi)^3f_c(x,{\bf q}|{\bf b})$ satisfies the conservation law \begin{equation}
\partial_\mu\left(\rho_c u^\mu\right)=0
\end{equation}
with the initial density determined by the nuclear
geometry $\rho_c(x_0|{\bf b})=T_A({\bf x}_t)T_B({\bf x}_t-{\bf b})\cosh\eta/\tau_0 d\sigma^\text{NN}_{c\bar c}/ d\eta$, where
$T_{A,B}({\bf x}_t)=\int_{-\infty}^{+\infty}\rho_{A,B}(\vec{r}) dz$ are the thickness functions, and $d\sigma^\text{NN}_{c\bar c}/d\eta$ is
the charm quark rapidity distribution in p+p collisions.

For the regeneration rate $\beta$, we also considered the canonical effect which is shown to be important in explaining the suppression of strange mesons~\cite{ko}. When there are only few pairs or even less than one pair of charm quarks produced in an event, one need to consider
the canonical effect to guarantee the exact charm number conservation. Taking into account the fact that the charm and anti-charm quarks inside a pair are produced
at the same rapidity, we simply multiply the regeneration rate $\beta$ in a unit rapidity bin by a canonical enhancement factor~\cite{liu}
\begin{equation}
\label{canonical}
C_{c\bar c}=1+1/(dN_{c\bar c}/dy).
\end{equation}

To take into account the relativistic effect on the dissociation cross section to avoid the divergence in the
regeneration cross section, we should replace the charmonium binding energy by the gluon threshold energy in the calculations of $\alpha$ and $\beta$~\cite{polleri}.

In the hadron phase of the fireball with temperature $T<T_c$, there are many effective
models that can be used to calculate the inelastic cross sections between charmonia and
hadrons~\cite{barnes}. For $J/\psi$ the dissociation cross section is about a few mb
which is comparable with the gluon dissociation cross section. However, considering that the hadron
phase appears in the later evolution of the fireball, the ingredient density of the system
is much more dilute in comparison with the early hot and dense period~\cite{tang}. Taking, for instance, the regeneration processes $c+\bar c \to g+J/\psi$ in quark matter and $D+\bar D^*\to \pi +J/\psi$ in hadron matter, the density ratio between charm quarks at initial temperature $T_0=310$ MeV and $D$ mesons at critical temperature $T_c=165$ MeV is around $30$. Considering further the life time of the quark matter $\sim 6$ fm/c and the life time of the hadron matter $\sim 2$ fm/c calculated from the hydrodynamics in Section \ref{s2}, we neglect the
charmonium production and suppression in hadron gas, to simplify the numerical calculations. Note that, the suppression and regeneration in hadron gas may become important for excited charmonium states~\cite{du}.

The transport equation can be solved analytically with the explicit solution~\cite{tang,liu4}
\begin{eqnarray}
\label{solution}
f_\Psi\left({\bf p}_t,y,{\bf
x}_t,\eta,\tau\right)&=&f_\Psi\left({\bf p}_t,y,{\bf
X}_t(\tau_0),H(\tau_0),\tau_0\right)\nonumber\\
&\times& e^{-\int^{\tau}_{\tau_0}{d\tau'\over \Delta(\tau')}
\alpha_\Psi\left({\bf p}_t,y,{\bf X}_t(\tau'),H(\tau'),\tau'\right)}\nonumber\\
&+&\int^{\tau}_{\tau_0}{d\tau'\over \Delta(\tau')} \beta_\Psi\left({\bf
p}_t,y,{\bf
X}_t(\tau'),H(\tau'),\tau'\right)\nonumber\\
&\times& e^{-\int^{\tau}_{\tau'}{d\tau''\over \Delta(\tau'')}\alpha_\Psi\left({\bf
p}_t,y,{\bf
X}_t(\tau''),H(\tau''),\tau''\right)}
\end{eqnarray}
with
\begin{eqnarray}
\label{xh}
&& {\bf X}_t(\tau')={\bf x}_t-{\bf
v}_T\left[\tau\cosh(y-\eta)
-\tau'\Delta(\tau')\right],\nonumber\\
&& H(\tau')=y-\arcsin\left(\tau/\tau' \sinh(y-\eta)\right),\nonumber\\
&&
\Delta(\tau')=\sqrt{1+(\tau/\tau')^2 \sinh^2(y-\eta)}.
\end{eqnarray}
The first and second terms on the right-hand side of the solution
(\ref{solution}) indicate the contributions from the initial
production and continuous regeneration, respectively, and both
suffer from the gluon dissociation in the medium. Since the regeneration
happens in the deconfined phase, the regenerated quarkonia
would have probability to be dissociated again by the surrounding gluons.
The coordinate shifts ${\bf x}_t \to {\bf
X}_t$ and $\eta \to H$ in the solution (\ref{solution})
reflect the leakage effect in the transverse and longitudinal
directions.

For fixed-target nuclear collisions at $E_\text{lab}$=2.76 A TeV, the collision time for the two Pb nuclei
to pass through each other in the center of mass frame is $2R_{\text{Pb}}m_\text{N}/(\sqrt{s_\text{NN}}/2)\sim 0.35$ fm/c, which is compatible with the charmonium
formation time but shorter than the QGP formation time $\tau_0=0.6$ fm. Therefore, all the cold nuclear matter effects can
be reflected in the initial charmonium distribution $f_\Psi$ at time $\tau_0$. We take into account nuclear absorption, nuclear shadowing and Cronin effect.
The initial distribution in the solution (\ref{solution}) can be obtained from a superposition of p+p collisions, along with
the modifications from these cold nuclear matter effects.

The nuclear absorption is important in explaining the $J/\psi$ suppression in p+A and A+A collisions at low energies. It is due to the inelastic collision between the initially produced charmonia and the surrounding nucleons, and its effect on the charmonium surviving probability can be described by an effective absorption
cross section $\sigma_\text{abs}$. The value of $\sigma_\text{abs}$ is usually measured in p+A collisions and is several mb at SPS energy. Since the nuclear absorption becomes weaker at higher colliding energy due to the shorter collision time\cite{capella,lourenco}, we take $\sigma_\text{abs}$=2 mb at $E_\text{lab}$=2.76 A TeV~\cite{lourenco} and the nuclear absorption factor
\begin{equation}
\label{abs}
S_\text{abs}=e^{-\sigma_\text{abs}\left(\int^{\infty}_{z_A}\rho(z,{\bf x_t}) dz + \int^{z_B}_{-\infty}\rho(z,{\bf x_t-b}) dz\right)}.
\end{equation}

The Cronin effect broadens the momentum distribution of the initially produced charmonia in heavy ion collisions~\cite{tang}. In p+A and A+A collisions, the incoming partons (both gluons and quarks) experience multiple scatterings with surrounding nucleons via soft gluon exchanges.  The initial scatterings lead to an additional
transverse momentum broadening of partons which is then inherited by produced hadrons~\cite{esumi}. Since the Cronin effect is caused
by soft interactions, rigorous calculations for the effect are not available. However, the effect is often treated as a random motion.
Inspired from a random-walk picture, we take a Gaussian smearing~\cite{zhao,liu2} for the modified transverse momentum distribution
\begin{equation}
\label{cronin}
\overline
f^\text{NN}_\Psi({\bf x},{\bf p},z_A,z_B|{\bf b})={1\over \pi a_{gN} l} \int
d^2{\bf p}_t' e^{-{\bf p}_t^{'2}\over a_{gN} l}f^\text{NN}_\Psi(|{\bf
p}_t-{\bf p}_t'|,p_z)S_\text{abs},
\end{equation}
where
\begin{equation}
l({\bf x},z_A,z_B|{\bf b})=\frac{1}{\rho} \left(\int_{-\infty}^{z_A}\rho(z,{\bf x_t}) dz + \int_{z_B}^{+\infty}\rho(z,{\bf x_t-b}) dz\right)
\label{ll}
\end{equation}
is the path length of the two initial gluons in nuclei before fusing into a charmonium at ${\bf x}$, $z_A$ and $z_B$, $a_{gN}$ is the averaged charmonium  transverse
momentum square gained from the gluon scattering with a unit of length of nucleons,
and $f^\text{NN}_\Psi({\bf p})$ is the charmonium momentum distribution in a free p+p collision. The Cronin parameter
$a_{gN}$ is usually extracted from corresponding p+A collisions.
Considering the absence of p+A
collision data at $\sqrt{s_\text{NN}}$= 72 GeV, we take $a_{gN}$=0.085 (GeV/c)$^2$/fm from some empirical
estimations~\cite{thews,wang2,vogt}. As a comparison, for collisions at
SPS ($\sqrt{s_\text{NN}} \sim 20$ GeV) and  RHIC ($\sqrt{s_\text{NN}}=200$ GeV) we take $a_{gN}=0.075$~\cite{zhu}  and
0.1~\cite{liu3} (GeV/c)$^2$/fm, respectively.

Assuming that the emitted gluon in the gluon fusion process $g+g\to
\Psi+g$ is soft in comparison with the initial gluons and the
produced charmonium and can be neglected in kinematics, the charmonium production becomes a $2\to 1$ process approximately, and the longitudinal
momentum fractions of the two initial gluons are
calculated from the momentum conservation,
\begin{equation}
\label{x}
x_{1,2}={\sqrt{m_\Psi^2+p_t^2}\over \sqrt{s_\text{NN}}} e^{\pm y}.
\end{equation}
The free distribution
$f_\Psi^\text{NN}({\bf p})$ can
be obtained by integrating the elementary partonic process,
\begin{equation}
\label{fg}
{d\sigma_\Psi^\text{NN}\over dp_tdy}= \int dy_g x_1 x_2 f_g(x_1,\mu_F)
f_g(x_2,\mu_F) {d\sigma_{gg\to\Psi g}\over d\hat t},
\end{equation}
where $f_g(x,\mu_F)$ is the gluon distribution in a free proton, $y_g$ the emitted
gluon rapidity, $d\sigma_{gg\to\Psi g}/ d\hat t$ the charmonium momentum distribution produced from a gluon fusion
process, and $\mu_F$ the factorization scale of the fusion process.

Now we consider
the shadowing effect. The distribution function $\overline
f_i(x,\mu_F)$ for parton $i$ in a nucleus differs from a superposition of the
distribution $f_i(x,\mu_F)$ in a free nucleon. The nuclear shadowing
can be described by the modification factor $R_i=\overline f_i/(Af_i)$.
To account for the spatial dependence of the shadowing in a finite
nucleus, one assumes that the inhomogeneous shadowing is
proportional to the parton path length through the nucleus~\cite{klein},
which amounts to consider the coherent interaction of the incident
parton with all the target partons along its path length. Therefore,
we replace the homogeneous modification factor $R_i(x,\mu_F)$ by an
inhomogeneous one~\cite{vogt2}
\begin{equation}
 {\cal R}_i(x,\mu_F,{\bf x}_t)=1+A\left(R_i(x,\mu_F)-1\right)T_A({\bf x}_t)/T_{AB}(0)
\end{equation}
with the definition $T_{AB}({\bf b})=\int d^2{\bf x}_t T_A({\bf x}_t) T_B({\bf x}_t-{\bf b})$.
We employ in the following the EKS98 package~\cite{eks} to evaluate the homogeneous
ratio $R_i$, and the factorization scale is taken as
$\mu_F=\sqrt{m_\Psi^2+p_t^2}$.

Replacing the free distribution $f_g$ in (\ref{fg}) by the modified
distribution $\overline f_g=Af_g{\cal R}_g$ and then taking into account the Cronin
effect (\ref{cronin}), we finally get the initial charmonium distribution for the
solution (\ref{solution}),
\begin{eqnarray}
\label{initial}
f_\Psi(x_0,{\bf p}|{\bf b})&=&{(2\pi)^3\over E_t\tau_0}\int dz_Adz_B\rho_A({\bf x}_t,z_A)\rho_B({\bf x}_t,z_B)\nonumber\\
&\times&{\cal R}_g(x_1,\mu_F,{\bf x}_t){\cal R}_g(x_2,\mu_F,{\bf x}_t-{\bf b})\nonumber\\
&\times& \overline f_\Psi^\text{NN}({\bf x},{\bf p},z_A,z_B|{\bf b})S_{abs}.
\end{eqnarray}
Now the only thing left is the distribution $f_\Psi^\text{NN}$ in a free p+p collision which
can be fixed by experimental data or some model simulations.

\section{Numerical Results}
\label{s4}

The beam energy $E_\text{lab}$= 2.76 A TeV in fixed target experiments corresponds to a colliding energy $\sqrt{s_\text{NN}}$=72 GeV, and the rapidity in the center-of-mass frame is boosted in the laboratory frame with a rapidity shift $\Delta y=\tanh^{-1}\beta_\text{cms}= 4.3$.
Let us first focus on the central rapidity region around $y_\text{cms}=$ 0 in the center-of mass frame, which corresponds to $y_\text{lab}=$ 4.3
in the laboratory frame. The centrality and momentum dependent anti-shadowing for initially produced charmonia is reflected in the inhomogeneous modification factor ${\cal R}_g$ for gluons.
The longitudinal momentum fractions
are $x_{1,2}=\sqrt{m^2_{\Psi}+p^2_t}/\sqrt{s_\text{NN}}\sim 0.05$ for the two gluons, which is located at the strong anti-shadowing region~\cite{dias} by some parametrization
of parton distribution shadowing like EKS98~\cite{eks}, EPS08~\cite{eps08} and EPS09~\cite{eps09}. The anti-shadowing changes not only the gluon distribution but also the charm quark production cross section used in the regeneration. For the process $g+g\to c+\bar c$, the anti-shadowing for gluons leads to an anti-shadowing factor $\sim ({\cal R}_g)^2$ for the cross section. Considering that in peripheral collisions the regeneration is weak and its contribution is not remarkably affected by the anti-shadowing, we take a centrality averaged anti-shadowing factor for the cross section to simplify the numerical calculation for regeneration. Estimated from the EKS98 evolution~\cite{eks}, we take a $20\%$ enhancement of the charm quark production cross section compared to free p+p collisions. From FONLL calculation~\cite{fonll},
the upper limit for $d\sigma_{c\bar c}^\text{NN}/dy$ is 0.047 mb at $\sqrt{s_\text{NN}}$=62.4 GeV. Note that the experimental data for charm quark
cross section in free p+p collisions are close to the upper limit of perturbative calculation, we take $d\sigma_{c\bar c}^\text{NN}/dy=0.05$ mb
at $\sqrt{s_\text{NN}}$=72 GeV. After taking into account the anti-shadowing effect in A+A collisions, it becomes 0.06 mb.
For p+p collisions,
we assume a constant hidden to open charm ratio $(d\sigma_{\Psi}/dy)/(d\sigma_{c\bar c}/dy)$=const at any colliding energy. From the ratio extracted from the RHIC data~\cite{adare}, we have $d\sigma_{J/\psi}/dy$=0.35 $\mu b$ at $\sqrt{s_\text{NN}}$=72 GeV. The transverse momentum distribution for $J/\psi$ in free p+p collisions
can be simulated by PYTHIA~\cite{pythia} and the mean transverse momentum square is $\langle p_t^2\rangle_\text{pp}=2.7$ (GeV/c)$^2$.

\begin{figure}[ht]
\centering
\includegraphics[width=0.45\textwidth]{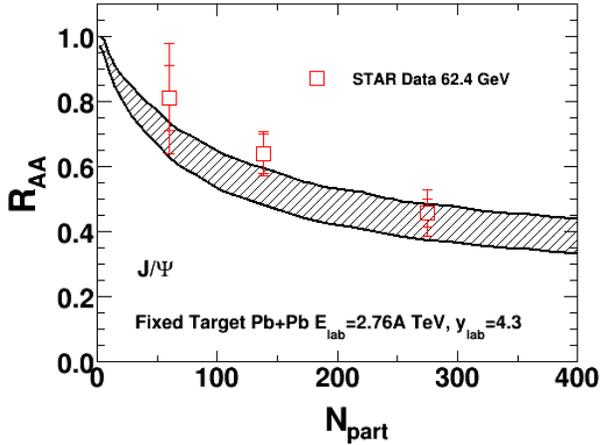}
\caption{(Color online) The centrality dependence of the $J/\psi$ nuclear modification factor $R_{AA}$ at very forward rapidity $y_\text{lab}=4.3$ ($y_\text{cms}$=0) in Pb+Pb collisions at LHC beam energy $E_\text{lab}$=2.76 A TeV. The hatched band is the model result with the upper and lower borders corresponding to the calculations with and without anti-shadowing effect. The RHIC data~\cite{rhic2} are for Au+Au collisions at $y_\text{cms}$=0. }
\label{fig1}
\end{figure}
Fig.\ref{fig1} shows our calculated centrality dependence of $J/\psi$ nuclear modification factor $R_{AA}=N_\Psi^{AA}/\left(N_\text{coll}N_\Psi^{pp}\right)$ in Pb+Pb collisions at LHC beam energy $E_\text{lab}$=2.76 A TeV in laboratory frame ($\sqrt{s_\text{NN}}$=72 GeV in center-of-mass frame) at forward rapidity $y_\text{lab}=4.3$ (central rapidity $y_\text{cms}$=0), where $N_\Psi^{pp}$ and $N_\Psi^{AA}$ are charmonium yields in p+p and A+A collisions, and $N_\text{coll}$ and $N_\text{part}$ are numbers of binary collisions and participants. For comparison, we show also the RHIC data at $\sqrt {s_{NN}}=62.4$ GeV~\cite{rhic2} at central rapidity. Since the shadowing/anti-shadowing effect is still an open question, and its degree depends strongly on the models we used, we show in Fig.\ref{fig1} two calculations for the total $J/\psi$ $R_{AA}$ in Pb+Pb collisions at $\sqrt{s_\text{NN}}$=72 GeV, one is with the above discussed anti-shadowing, and the other is without anti-shadowing. The hatched band is due to this uncertainty in the anti-shadowing. With increasing collision centrality, the initial contribution drops down, while the regeneration goes up. The canonical effect is important
in peripheral collisions where the number of charm quark pairs is less than one and the inclusion of the canonical effect
enhances sizeably the charmonium yield. In most central collisions, the regeneration can contribute about $25\%$ to the total charmonium yield. The anti-shadowing at very forward rapidity in the laboratory frame (central rapidity in the center-of-mass frame) enhances the charm quark cross section and in turn the initial charmonium yield by a factor of 1.2. As a consequence, the enhancement factor for the regenerated charmonium number is $1.2^2=1.44$ which leads to a strong charmonium enhancement! If we do not consider the anti-shadowing effect on the charmonium regeneration
and initial production, the total $R_{AA}$ is significantly reduced.

\begin{figure}[ht]
\centering
\includegraphics[width=0.45\textwidth]{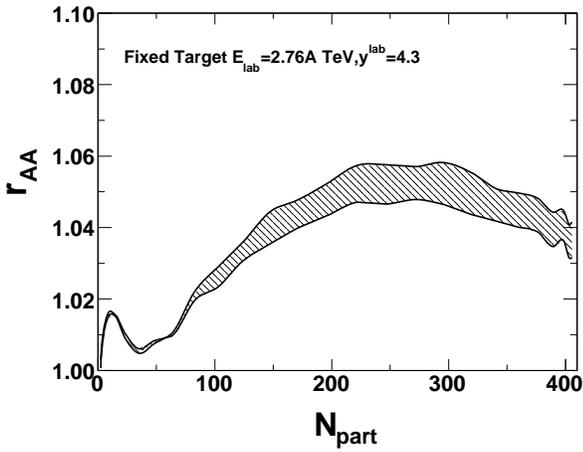}
\caption{(Color online) The centrality dependence of the $J/\psi$ nuclear modification factor $r_{AA}$ at forward rapidity $y_\text{lab}$=4.3 $(y_\text{cms}$=0) in Pb+Pb collisions at LHC beam energy $E_\text{lab}$=2.76 A TeV. The upper and lower borders of the band correspond to the calculations with and without anti-shadowing effect. }
\label{fig2}
\end{figure}
To see more clearly the charmonium production mechanism, we turn to the transverse momentum information. In Fig.\ref{fig2}
we show the $J/\psi$ nuclear modification factor~\cite{zhou}
\begin{equation}
r_{AA}={\langle p_t^2\rangle_{AA}\over \langle p_t^2\rangle_{pp}}
\end{equation}
in Pb+Pb collisions at beam energy $E_{\text{lab}}$=2.76 A TeV, where $\langle p_t^2\rangle_{AA}$ and $\langle p_t^2\rangle_{pp}$ are averaged $J/\psi$ transverse momentum square in Pb+Pb and p+p collisions at very forward rapidity $y_\text{lab}$=4.3. If we neglect the contribution from the regeneration and consider only the initial production, the ratio $r_{AA}$ goes up monotonously with centrality due to the Cronin effect and leakage effect~\cite{zhou}. The inclusion of regeneration (upper border of the band) remarkably reduces the averaged transverse momentum, because the regenerated charmonia possess a soft momentum distribution induced by
the charm quark energy loss. Since the degree of regeneration increases with centrality, the increased soft component leads to a decreasing $r_{AA}$ in most central collisions. The canonical effect can reduce the $r_{AA}$ further, since it enhances the regeneration especially in peripheral collisions. However, we should note that, the assumption of charm quark thermalization indicates a full energy loss and it may not be reached in peripheral and semi-central collisions at beam energy $E_\text{lab}$=2.76 A TeV. When we switch off the anti-shadowing (lower border of the band), both the hard component controlled by the initial production and the soft component dominated by the regeneration would be reduced. Considering that the enhancement factor resulted from the anti-shadowing is $1.2$ for the initial production but $1.2^2$ for the regeneration, the stronger anti-shadowing in the soft component leads to the slight difference between with and without considering the anti-shadowing, shown in Fig.\ref{fig2}. It is obvious that compared to the nuclear modification
factor $R_{AA}$ for the yield, the modification factor $r_{AA}$ for the transverse momentum is less sensitive to the shadowing effect~\cite{zhou}.

\begin{figure}[ht]
\centering
\includegraphics[width=0.45\textwidth]{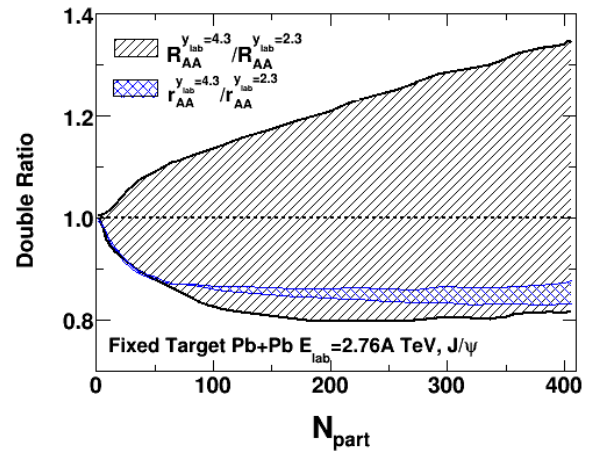}
\caption{(Color online) The centrality dependence of the double ratios $R_{AA}^{y_\text{lab}=4.3}/R_{AA}^{y_\text{lab}=2.3}$ and $r_{AA}^{y_\text{lab}=4.3}/r_{AA}^{y_\text{lab}=2.3}$ for $J/\psi$ yield and transverse momentum in Pb+Pb collisions at LHC beam energy $E_\text{lab}$=2.76 A TeV. The upper and lower borders of the two bands correspond to the calculations with and without shadowing and anti-shadowing effects. }
\label{fig3}
\end{figure}
From the simulations of parton distributions in cold nuclear matter~\cite{eks,eps08,eps09}, the nuclear shadowing region is located at very small $x$. In the following we consider the shadowing and see its difference from the anti-shadowing in $J/\psi$ $R_{AA}$ and $r_{AA}$ in fixed-target Pb+Pb collisions. The maximum $J/\psi$ rapidity in the center-of-mass frame is $y_\text{cms}^\text{max}=\cosh^{-1}\left[\sqrt{s_\text{NN}}/\left(2m_{J/\psi}\right)\right]$=3.13 at $\sqrt{s_\text{NN}}$=72 GeV.
Considering the expected amount of measured events, we focus on the backward rapidity region around $y_\text{cms}=-2$ which corresponds
to the less forward rapidity $y_\text{lab}=\Delta y+y_\text{cms}=4.3-2=2.3$ in laboratory frame. From the kinematics, the momentum fractions for the two gluons involved in the gluon fusion process are $x_1=(\sqrt{m_\Psi^2+p_t^2}/\sqrt{s_\text{NN}})e^2 = 0.35$ and $x_2=(\sqrt{m_\Psi^2+p_t^2}/\sqrt{s_\text{NN}})e^{-2} = 0.006$. One is located in the EMC region and the other in the shadowing region~\cite{eks,eps08,eps09}, leading to a reduction of $15\%$
for the charm quark production cross section from EKS98 evolution~\cite{eks} ($20\%$ from EPS09 NLO evolution~\cite{eps09}). Taking the same
ratio of charm quark cross section between $y_\text{cms}=-2$ and $y_\text{cms}=0$ calculated from FONLL~\cite{fonll} and including the
$15\%$ shadowing reduction, we obtain $d\sigma_{c\bar c}^\text{NN}/dy$=0.01 mb at $y_\text{cms}$=-2. For the medium evolution at
this backward rapidity region, we initialize the entropy density to be half of that at central rapidity~\cite{shen,na50} which leads to a maximum temperature of $T_0$=245 MeV.
Fig.\ref{fig3} shows the two double ratios $R_{AA}^{y_\text{lab}=4.3}/R_{AA}^{y_\text{lab}=2.3}$ and $r_{AA}^{y_\text{lab}=4.3}/r_{AA}^{y_\text{lab}=2.3}$ of $J/\psi$, the upper and lower borders of the two bands correspond to the calculations with and without considering the nuclear shadowing and anti-shadowing. While the double ratio for the transverse momentum is not sensitive to the shadowing and anti-shadowing, as we discussed above, the strong anti-shadowing at $y_\text{lab}=4.3$ and shadowing at $y_\text{lab}=2.3$ leads to a strong enhancement of the double ratio for the yield. Without considering the shadowing and anti-shadowing, the stronger charmonium suppression in the hotter medium at $y_\text{lab}=4.3$ ($T_0$=310 MeV) compared with the weaker suppression in the relatively colder medium at $y_\text{lab}=2.3$ ($T_0$=245 MeV) makes the double ratio less than unit. However, the inclusion of the yield enhancement due to the anti-shadowing at $y_\text{lab}=4.3$ and the yield suppression due to the shadowing at $y_\text{lab}=2.3$ changes significantly the behavior of the double ratio, it becomes larger than unit and can reach 1.3 in most central collisions. Note that the rapidity dependent shadowing effect was used to qualitatively interpret the stronger suppression at forward rapidity than that at midrapidity in Au+Au collisions at RHIC~\cite{frawley,ferreiro}.

\section{Summary}
\label{s5}

We investigated with a transport approach the charmonium production in fixed-target Pb+Pb collisions at LHC beam energy $E_\text{lab}$=2.76 A TeV. We focused on the rapidity dependent shadowing effect on the nuclear modification factors for the charmonium yield and transverse momentum. While the averaged transverse momentum is not sensitive to the shadowing effect, the anti-shadowing leads to a strong yield enhancement at very forward rapidity $y_\text{lab}\simeq$ 4, and the shadowing results in a strong yield suppression at less forward rapidity $y_\text{lab}\simeq$2. The double ratio between the nuclear modification factors $R_{AA}$ in the two rapidity regions amplifies the shadowing effect, it is larger than unit and can reach 1.3 in most central collisions.

From the model studies on gluon distribution in nuclei, see for instance Refs.~\cite{eks,dias,eps08,eps09}, there are large uncertainties in the domain of large $x\ (>0.1)$, which is probably due to the unknown EMC effect. From our calculation here, the double ratio of the nuclear modification factor for $J/\psi$ yield is very sensitive to the gluon shadowing effect in different $x$ region. A precise measurement of the ratio may provide a sensitive probe to the gluon distribution.
\\ \\
{\bf Acknowledgement:} The work is supported by the NSFC under grant No.11335005 and the MOST under grant Nos.2013CB922000 and 2014CB845400.

\end{document}